\documentclass[11pt,twoside]{article}
\usepackage{./asp2014}

\aspSuppressVolSlug
\resetcounters

\begin{document}

\title{The Montreal White Dwarf Database: a Tool for the Community}
\author{P. Dufour, S. Blouin, S. Coutu, M. Fortin-Archambault, C. Thibeault, P. Bergeron and G. Fontaine \affil{Universit\'e
    de Montr\'eal, Montr\'eal, Qu\'ebec, Canada;
    \email{dufourpa@astro.umontreal.ca}}}

% This section is for ADS Processing.  There must be one line per author.
\paperauthor{Sample~Author1}{Author1Email@email.edu}{ORCID_Or_Blank}{Author1 Institution}{Author1 Department}{City}{State/Province}{Postal Code}{Country}
\paperauthor{Sample~Author2}{Author2Email@email.edu}{ORCID_Or_Blank}{Author2 Institution}{Author2 Department}{City}{State/Province}{Postal Code}{Country}
\paperauthor{Sample~Author3}{Author3Email@email.edu}{ORCID_Or_Blank}{Author3 Institution}{Author3 Department}{City}{State/Province}{Postal Code}{Country}

\begin{abstract}
We present the "Montreal White Dwarf
Database"\footnote{\url{www.montrealwhitedwarfdatabase.org}} (MWDD), an
accessible database with sortable/filterable table and interactive
plots that will, when fully completed, allow the community to explore
the physical properties of all white dwarfs ever analyzed by the
Montreal group, as well as display data and analyses from the
literature. We present its current capability and show how it will
continuously be updated to instantly reflect improvements made on both
the theoretical and observational fronts.
\end{abstract}

\section{Introduction}
The last decade or so has seen a dramatic increase in the number of
spectroscopically identified white dwarf stars, going from about 2500
at the beginning of the millennium to around 30,000 at the time of
this writing. For many years, one of the most important place to look
for information about white dwarf stars has been the George P. McCook
and Edward M.Sion White Dwarf Catalog ({\url
  www.astronomy.villanova.edu/WDcatalog/}). However, this catalog
currently contains "only" 14294 objects (it was last updated in 2013),
leaving many new SDSS white dwarfs behind. Moreover, this invaluable
resource contains very little information on the physical parameters
published in the literature.

As a result of the huge amount of data and analyses now available,
keeping an eye on the big picture has become increasingly
difficult. Many basic questions now require a tremendous amount of
work just to get updated and obtain accurate answers. Questions such as: How
complete is the census of white dwarfs within a given distance of the
Sun? How many have metals, are magnetic, are in binary systems, or are
He-rich? What fraction of these are there as a function of effective
temperature?  What are the properties of any given population? What
are their mass distributions, luminosity functions, etc.? To answer
those questions, one usually needs to compile all the information from
the literature, double check it, look out for updates, and assess how
recent data/models change the picture. This is a very time-consuming
task that has to be repeated again and again as time moves
forward. Even the most comprehensive efforts are practically
out-of-date very soon after they are published \citep[for the local
  sample,][come to mind]{Holberg08,Giammichele12}.

A tool that would allow to easily get answers to those basic questions
would certainly be useful to all of us. This is why we decided to make
good use of the little surplus money from our organization of the
previous workshop in Montreal and give it back to the community by
creating this database. In what follows we present the main
characteristics of MWDD as well as our vision of it in the near
future.

\section{Database Description}

The Montreal White Dwarf Database aims to gather in one place all the
information and available data about the spectroscopically identified
white dwarf that have been discovered to this day. Interactive tables
and tools to easily make plots, histograms or display data have also
been implemented (see below). The structure and philosophy behind MWDD
was inspired in parts by other databases available to the exoplanets
community such as {\url http://www.openexoplanetcatalogue.com/} and
{\url http://exoplanets.org/}. 

We constructed a compilation of known white dwarf stars by simply
copying in spreadsheets (one .csv file per paper/source) all the
information (names, magnitudes, effective temperature, surface
gravity, abundances etc.) contained in the tables of well known papers
that analyzed large samples (we also included compilations from a few
review papers and websites). Since a star can be published under many
different names, each star is assigned a unique identifier via
Simbad. This allows us to automatically merge together as one entry in
the database objects entered under different names in different
papers. MWDD currently includes data compiled from 120 papers for a
total approaching 30,000 white dwarfs. As more papers are linked to
the database its degree of completeness will improve.

\subsection{Search Filters}

We implemented numerous filtering options to allow anyone to quickly
isolate certain types of white dwarfs of interest or do some
statistics. It is currently possible to filter by numerous parameters
such as coordinates, effective temperature, mass, distance and
magnitude as well as some important white dwarf characteristics such
as their spectral type, surface composition, variability or
magnetism. Plots, histograms, statistics and the table (see below) are
automatically updated to reflect the choices made in the search filter
section.

\articlefigure[width=1.\textwidth]{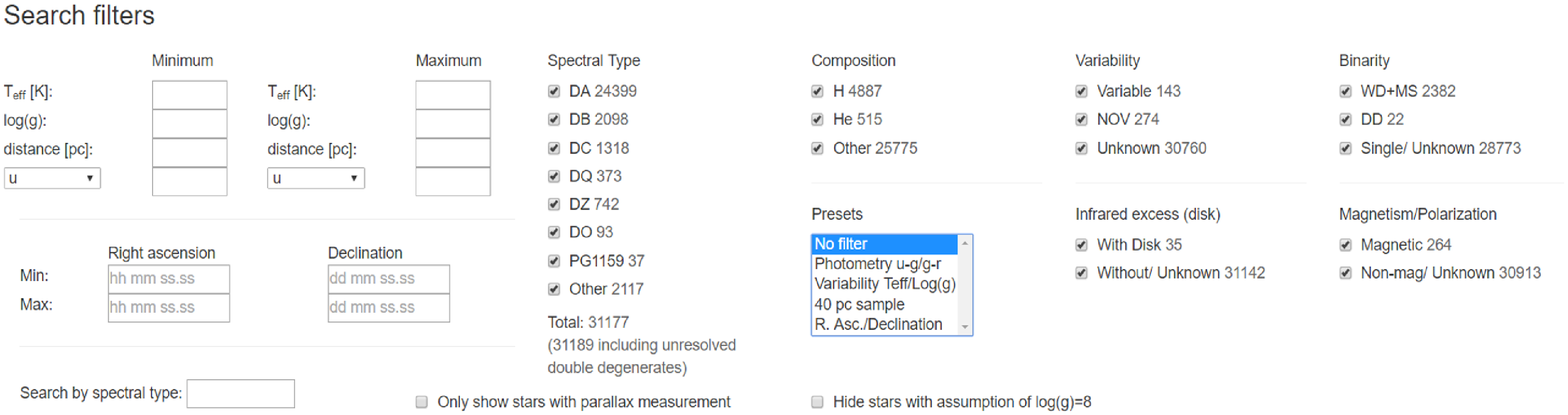}{ex_fig1}{Filter
  interface in MWDD.}

\subsection{Table}

All the properties compiled in the database can be selected and viewed
via the "Table" tab. Table columns are dragable and sortable. Clicking
on "Options" allow the user to select among a long list of available
parameters. One can also search by name, or enter a list of names. It
is also possible to export the data of all objects that pass the filters.

The values presented in the table are the last published ones (except
for a few sources which have been given lower priority). In the
future, a quality flag to use what we consider the best values will be
implemented. Meanwhile, we encourage users to go on the individual
white dwarf page in order to view the various literature values
included in the database and assess themselves which parameters are to
be trusted most.

\articlefigure[width=1.\textwidth]{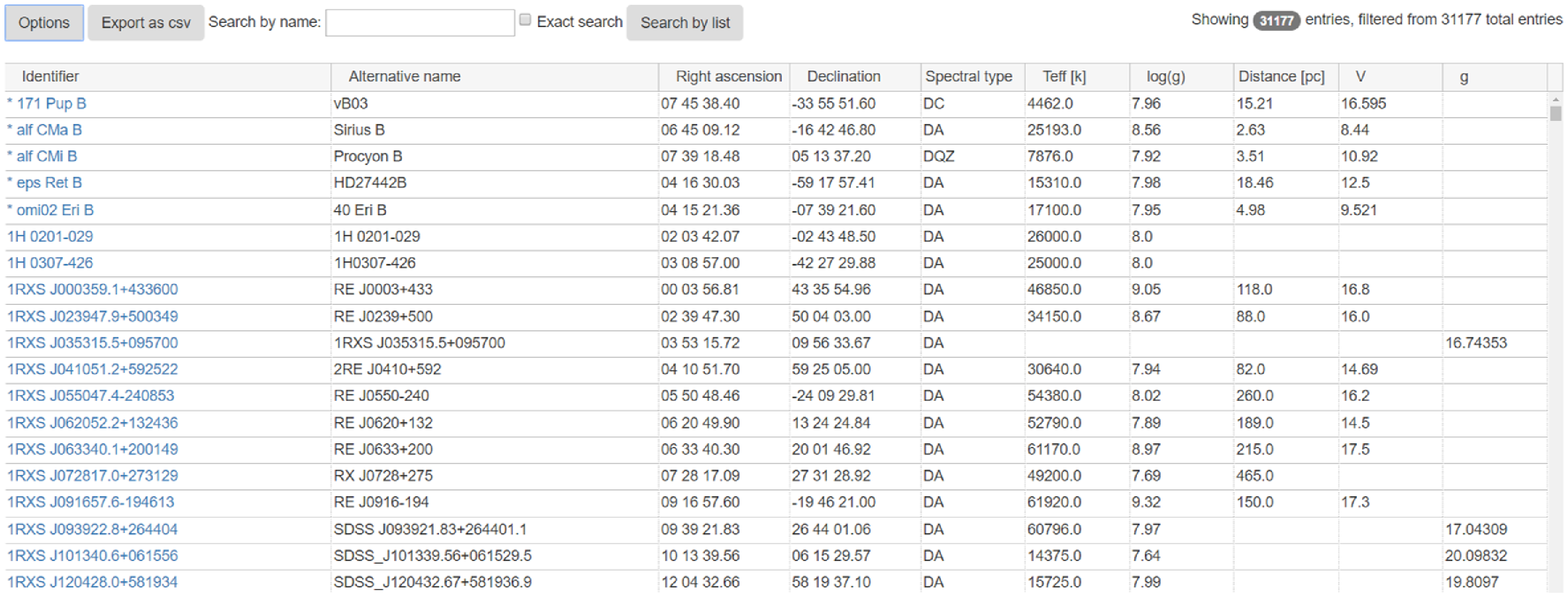}{ex_fig1}{An example of
  the table interface.}

\subsection{Plots, Histograms and Statistics}

JavaScript is used to make the numerous interactive plots and
histograms in the user interface. The "Scatter Plot" tab allows to plot
any physical quantities as a function of another. Only stars passing
the selected filters will appear. Moving the mouse on an object
highlights its basic atmospheric properties and clicking on it opens
an individual white dwarf page (see below). It is possible to export
the stars data present in a plot in order to make your own plot by
selecting "Show only stars visible in Scatter Plot" in the "Table" tab.

\articlefigure[width=1.\textwidth]{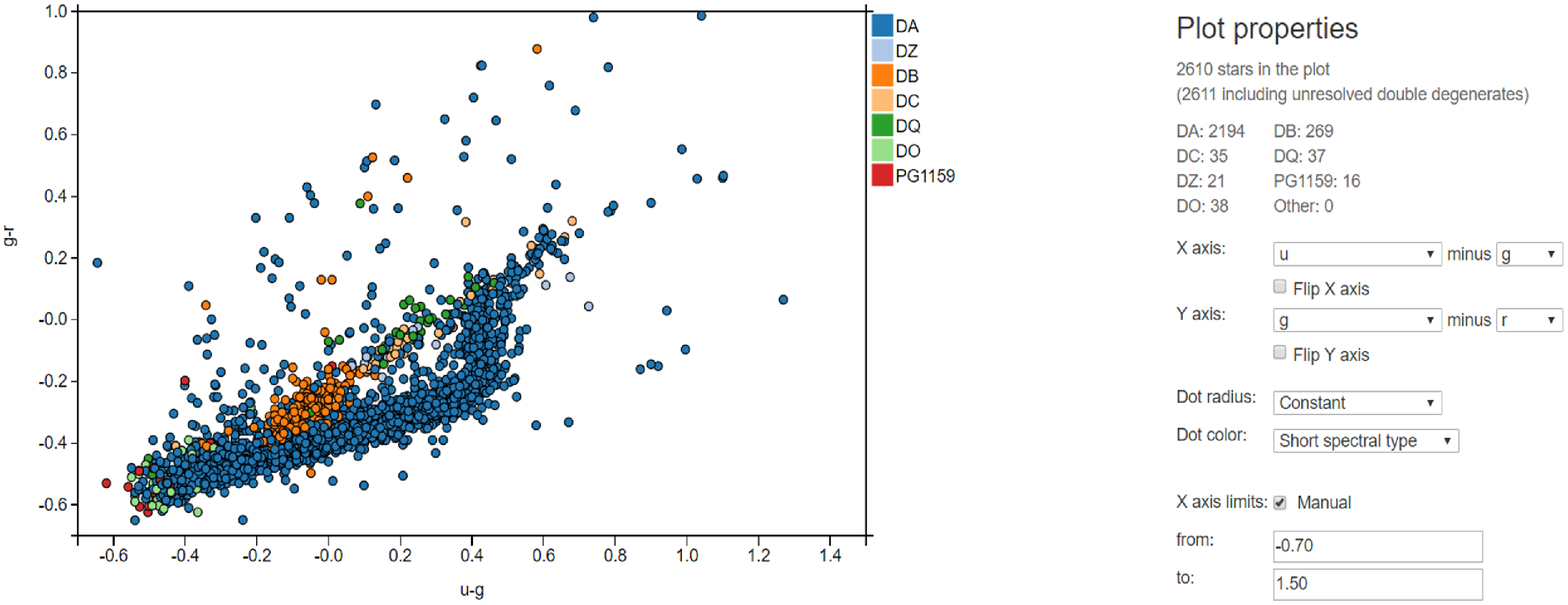}{ex_fig1}{u-g vs g-r
  diagram for white dwarfs brighter than g = 17.75.}

The "Histogram" tab allows, for example, to quickly get the mass
distribution of selected objects or distribution of any other variable
of interests. One can also easily get a measure of the current status
of our census of white dwarfs within a given distance from the Sun in
the "Cumulative Number" tab \citep[see][]{Holberg08}. For example, by
clicking on "40 pc sample" in the Presets options, one can see that we
estimate a 50.0$\%$ completeness within that distance. These numbers
are automatically updated as new papers are published and included in
the database, allowing the community to get an instant picture of our
neighborhood.
 
\articlefigure[width=1.\textwidth]{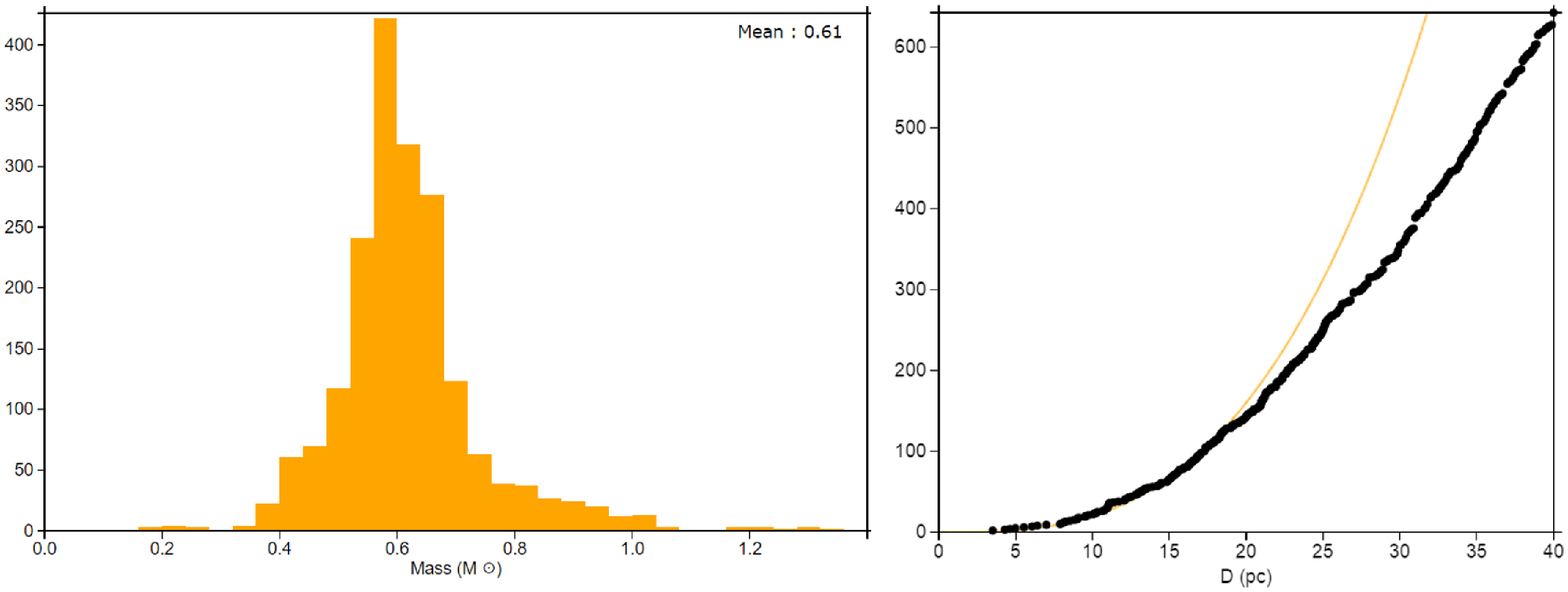}{ex_fig1}{Mass
  distribution of DA white dwarfs (left) and cumulative number as a
  function of distance for all white dwarfs within 40 pc (right). The
  straight line represents the expected number of white dwarfs for a
  uniform space density of 4.8$\times 10^{-3}$pc$^{-3}$.}

\subsection{Individual white dwarf page}

\articlefigure[width=1.\textwidth]{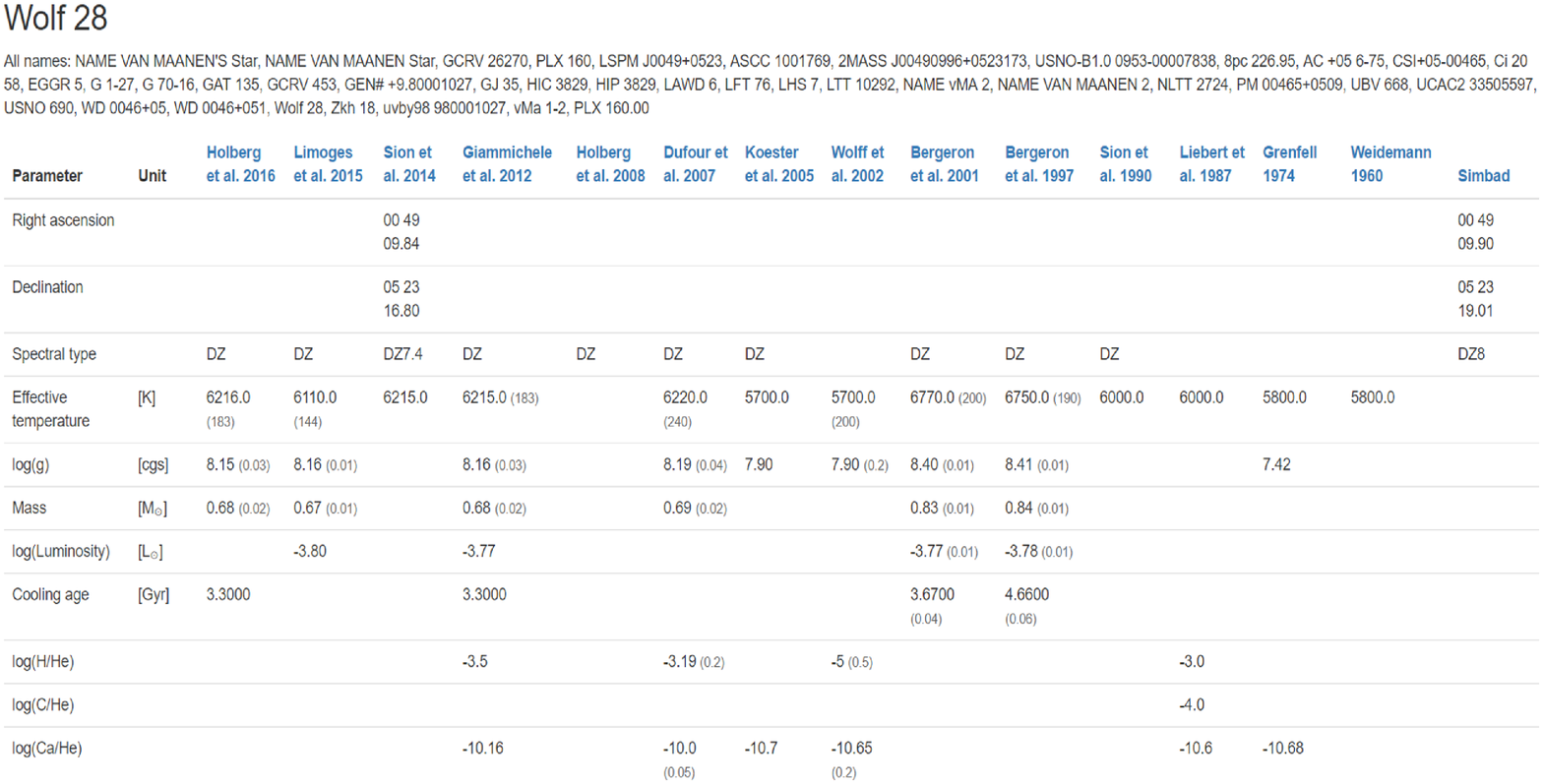}{ex_fig1}{An example of an
  individual white dwarf page (van Maanen 2).}

Each object in the database has an individual page (accessible via the
"Table" tab, or by clicking on an object in the "Scatter Plot"
tab). All the references included in the database that contain that
object are listed for an easy comparison between parameters from
different sources. Spectroscopic data, when available, can also be
visualized interactively. This includes all SDSS spectra, IUE archive
data, and many of the spectra secured by the Montreal group in the
last 3 decades. Optical spectra for more than 97 $\%$ of the stars
included in the database can currently be viewed and downloaded. We
hope MWDD eventually becomes a standard repository of spectra for
everyone to use. If you are in possession of worthy data and wish to
share them with the community, we will be glad to link them to the
database.

\subsection{Evolutionary models and diffusion timescales}

One of the most frequent request we routinely receive concerns the
white dwarf properties (mass, radius, cooling age etc.) for a given
surface gravity, temperature and composition. These numbers can now be
easily accessed directly on the MWDD webpage. For pedagogical purpose,
a cartoon of the white dwarf size relative to Earth also automatically
appears when the user enters log g/$T_{\rm eff}$. Another frequent
request concerns the diffusion timescale of heavy elements at the
surface of white dwarf stars. Some 25 years after \citet{Paquette86},
we improved significantly on their calculations of diffusion
coefficients by 1) using a much more accurate numerical scheme for
estimating the so-called collision integrals, and 2) introducing a
more physical prescription of the screening length used as the
independent variable in the evaluation of these coefficients. Details
of these improved calculations will be provided elsewhere. An
interpolation routine allows the users to enter log g and $T_{\rm
  eff}$ and obtain the settling timescales for all 27 elements from Li
to Cu in the periodic table (also downloadable as a plain text file).

\articlefigure[width=1.\textwidth]{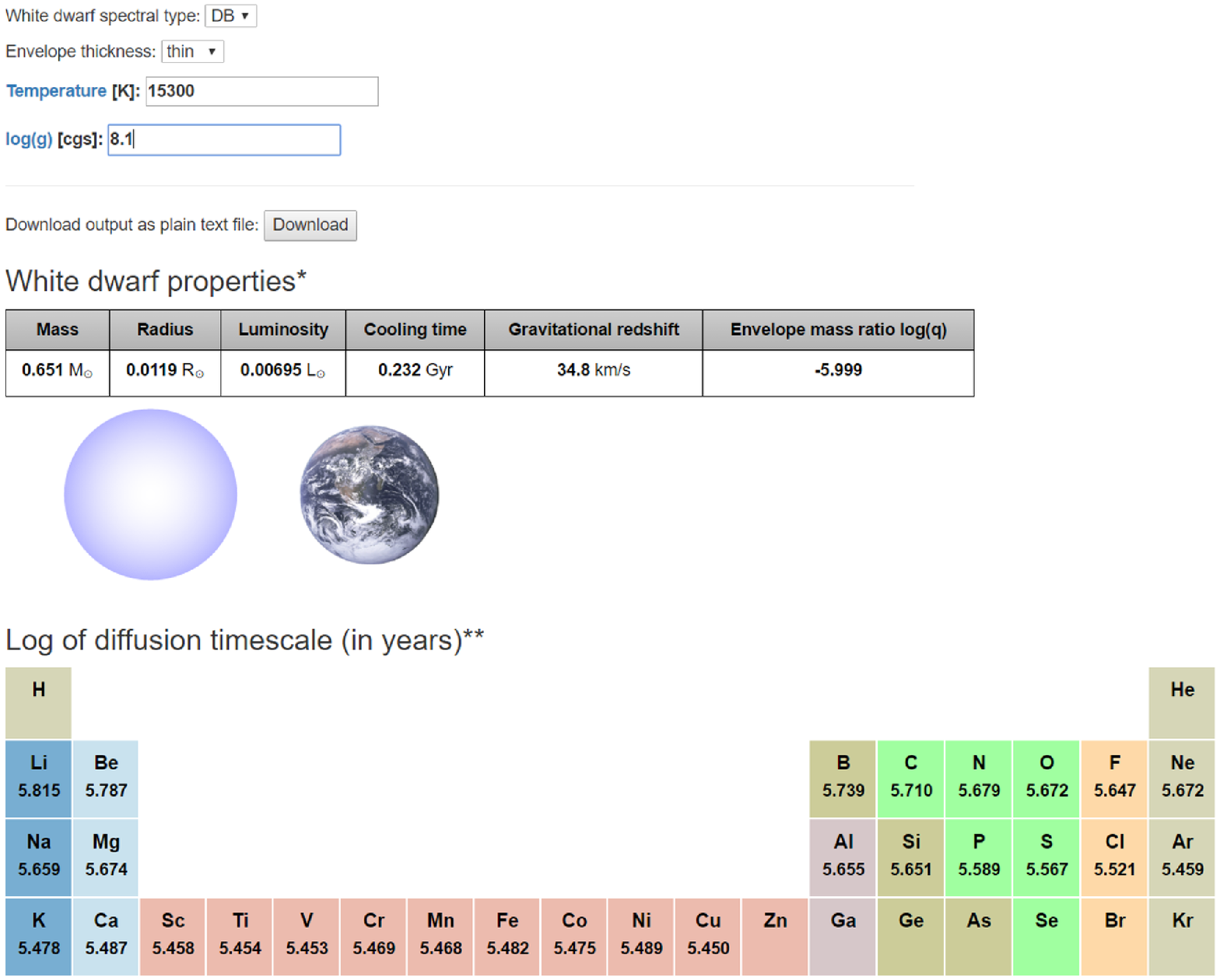}{ex_fig1}{White dwarf
  properties and diffusion timescales}

\subsection{What's next?}

The next step, among other things, to improve the usefulness of the
database to the community will be to add the possibility to make
luminosity functions and filter with S/N ratio, wavelength range
coverage and resolution of the spectroscopic observations. A comment
section to discuss individual object or general topics will also be
implemented. We also plan to include our own results from model
atmosphere fitting. In the near future, we will include
photometric/spectroscopic fits and distance estimate for every simple
objects present in the database (i.e. stars that can easily be fitted
with a standard grid such as single DA, DB, DC and DQ stars). The
homogeneity of the analysis should allow a much clearer view of the
properties of white dwarf stars in general. Moreover, once all the
data from the database will be directly linked with our fitting
routine, it will be much easier to update the parameters whenever new
data, or new grids become available. For example, when Gaia's
parallaxes become available, it will be possible to provide instantly
to the community the mass distribution of cool white dwarfs by simply
adding one single file to the database (white dwarf names +
parallaxes) and then running our fitting routines (which typically
takes a few seconds per objects). More complicated stars that stand
out will then be treated accordingly and included once we are
satisfied with the results.

We should stress that the database is not complete (it only includes
data found in a few selected papers) and should not replace in any way
a thorough search through the literature. It is nevertheless a good
place to start to quickly get information about any given star. While
we will periodically link new paper's data to the database, it is our
hope the members of the community will participate in our efforts to
make the database as complete as possible by submitting .csv files
(templates are available) for their own work or interesting papers not
included yet. We welcome contributions and corrections from all
professional astronomers and white dwarf aficionados.

If using MWDD was useful for your research, please cite this
manuscript and include an appropriate acknowledgment, a very simple
action that will help us tremendously in getting the funding necessary
to maintain/improve MWDD in the future.

\acknowledgements This work was supported in part by the NSERC Canada
and by the Fund FRQ-NT (Québec). The participants of the The 19th
European Workshop on White Dwarfs
(\url{http://craq-astro.ca/eurowd14/IMG_8148_modif.jpg}).

%\bibliography{editor}  % For BibTex

% For non-BibTex:

\end{document}